\documentclass{article}    

\def\ang{\AA}
\def\hbar{{\mathchar'26\mkern-9muh}}

\def\gapprox{\lower.4ex\hbox{$\;\buildrel >\over{\scriptstyle\sim}\;$}}
\def\lapprox{\lower.4ex\hbox{$\;\buildrel <\over{\scriptstyle\sim}\;$}}

\def\ref#1{\par\noindent\hangindent1cm {#1}}

\def\captio#1{\caption{\small {#1} \normalsize}}

\usepackage{graphics}
\usepackage{graphicx}
\usepackage{epsfig}
\usepackage{times}
\usepackage{makeidx}

\begin{document}           
\renewcommand{\topfraction}{0.95}
\renewcommand{\bottomfraction}{0.95}
\renewcommand{\textfraction}{0.05}
\renewcommand{\floatpagefraction}{0.95}
\renewcommand{\dbltopfraction}{0.95}
\renewcommand{\dblfloatpagefraction}{0.95}

\par\noindent
{\bf{Chaos and Complex Systems - CCS2010}}
\par\noindent
$3^{rd}$ International Interdisciplinary Chaos Symposium 
\par\noindent
Istanbul, Turkey, 21-24 May 2010.

\bigskip
\begin{center}
\LARGE{Self-Organized Criticality in Solar Physics and Astrophysics}
\end{center}

\medskip
\begin{center}
{\large Markus J. Aschwanden}     
\end{center}

\medskip
\begin{center}
{\it Solar and Astrophysics Laboratory, Lockheed Martin, Palo Alto, USA - 
e-mail: aschwanden@lmsal.com}
\end{center}

\begin{abstract}
The concept of ``self-organized criticality'' (SOC) has been introduced
by Bak, Tang, and Wiesenfeld (1987) to describe the statistics of avalanches
on the surface of a sandpile with a critical slope, which produces
a scale-free powerlaw size distribution of avalanches. In the meantime,
SOC behavior has been identified in many nonlinear dissipative systems 
that are driven to a critical state. On a most general level, SOC is the 
statistics of coherent nonlinear processes, in contrast to the Poisson 
statistics of incoherent random processes. The SOC concept has been applied
to laboratory experiments (of rice or sand piles), to human activities 
(population growth, language, economy, traffic jams, wars),
to biophysics, geophysics (earthquakes, landslides, forest fires),
magnetospheric physics, solar physics (flares), stellar physics
(flares, cataclysmic variables, accretion disks, black holes,
pulsar glitches, gamma ray bursts), and to galactic physics and
cosmology. 
\end{abstract}

\medskip
{Keywords: astrophysics - nonlinear dynamics - statistics -
 selforganized criticality - solar flares}

\section{Introduction}

Physical processes in our universe can be subdivided into incoherent 
and coherent processes. 
Examples of incoherent processes are fractional Brownian motion,
diffusion, collisional plasma processes, plasma heating, which
all are governed by random processes and thus can be described
by binomial and Poissonian statistics, which exhibit exponential-like 
or Gaussian-like distributions. In contrast, coherent processes 
overcome the threshold of the random noise background and grow
in a multiplicative way, such as avalanches, chain reactions, or
catastrophes, which all exhibit powerlaw-like size distributions.
The dychotomy of incoherent and coherent processes
is also reflected in the two types of linear and nonlinear processes.
Linear systems are characterized by a proportionality between the
input and output. Small disturbances cause small effects, and large
disturbances are required to produce a large effect. A hydropower plant,
for instance, produces electric energy that is proportional to the
water inflow rate and gravitational or kinetic energy of the water
that is feeding the turbine (Fig.1, top). Nonlinear processes, in
contrast, are governed by a multiplicative or amplification factor,
that leads to outputs of unpredictable magnitude, unlike linear systems. 
Small disturbances
can cause both small or large avalanches. Classical examples are
snow avalanches (Fig.~1, bottom), landslides, floodings, forest fires,
or earthquakes. 

\begin{figure}
\centerline{\includegraphics[width=1.0\textwidth]{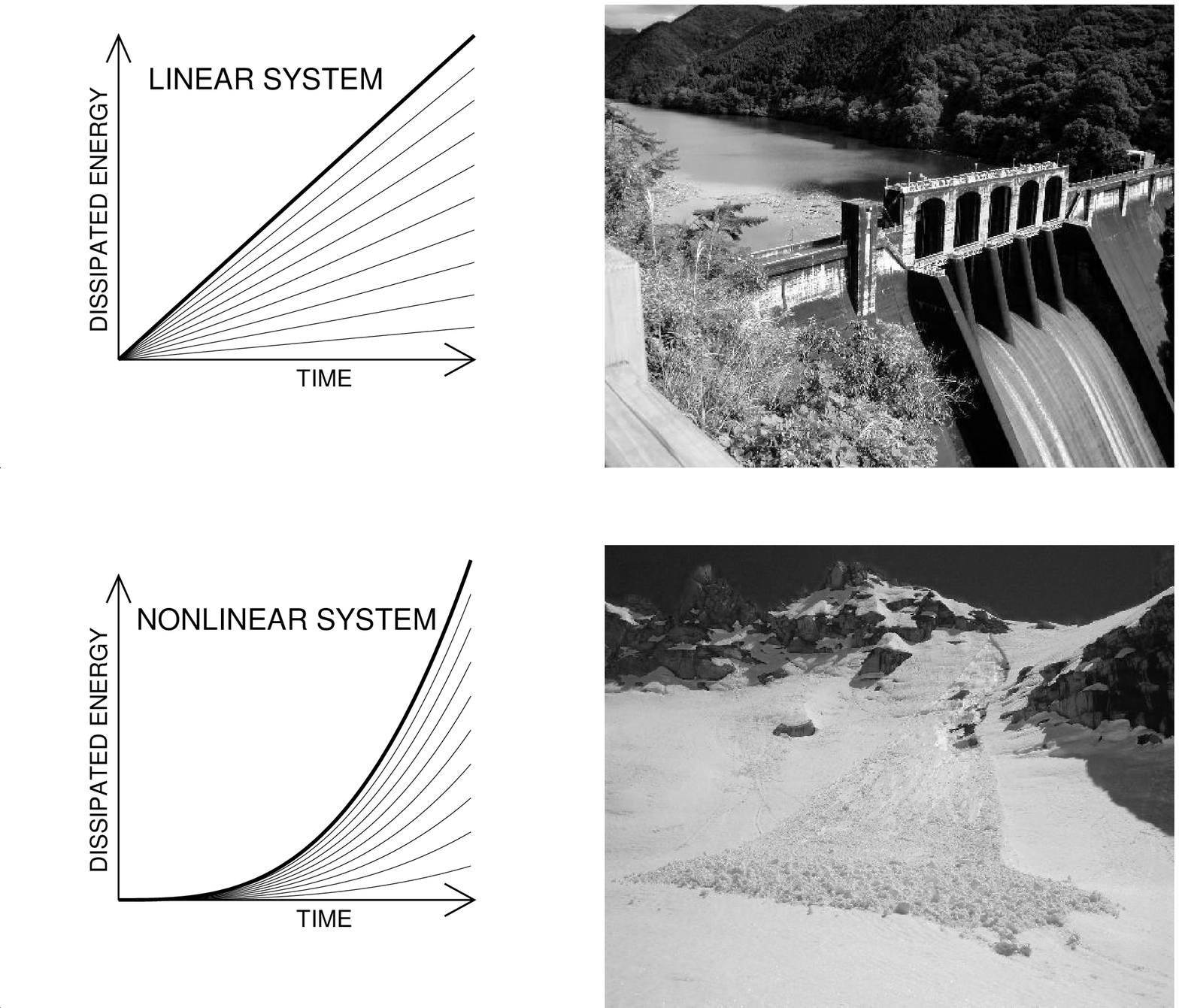}}
\captio{The output or dissipated energy in a linear system grows
linearly with time, for a constant input rate (top left), while
the output is highly unpredictable and not correlated with
the input rate in a nonlinear dissipative system (bottom left).
A practical example of a linear system is
a hydroelectric plant, where the produced electric energy is proportional
to the water input, as depicted with the water-storage dam at Yaotsu,
Gifu, Japan (top right). A classical example of a dissipative nonlinear system
is a snow avalanche, as shown in the large wet-snow avalanche at
Deadman Canyon in the Sierra Nevada range (bottom right).}
\end{figure}

\begin{figure}
\centerline{\includegraphics[width=1.0\textwidth]{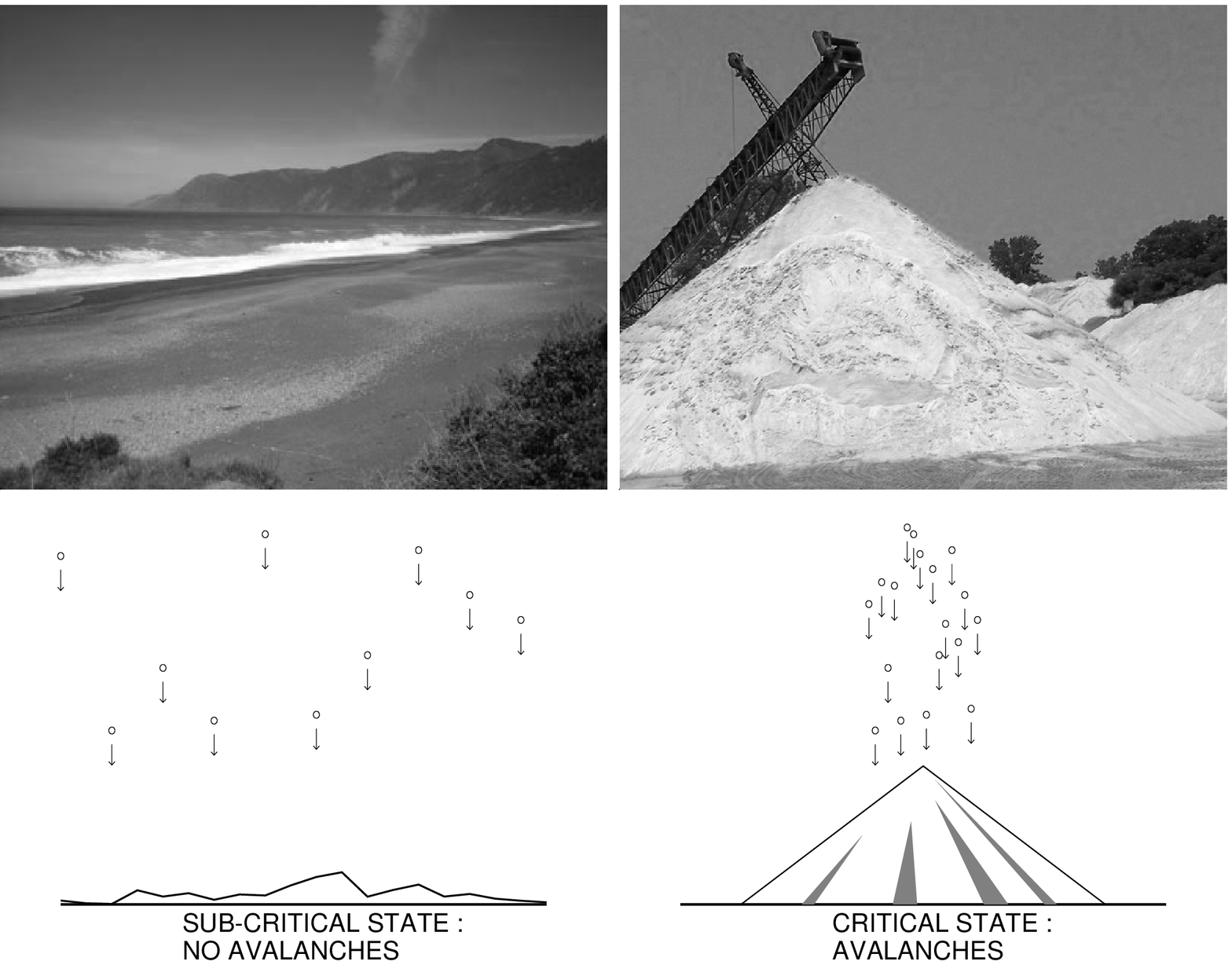}}
\captio{A static equilibrium produces no avalanche events (bottom left panel),
such as the flat sand beach in northern California (top left panel), while
randomly dripping sand onto a sandpile produces a state of self-organized
criticality where avalanches occur (bottom right panel), such as with the
conveyer belt of the Indian River Enterprises (top right panel).}
\end{figure}

A conceptual model to understand the basic nature of nonlinear processes
was introduced by Bak, Tang, and Wiesenfeld (1987) in terms of sandpile
avalanches. We can build a sandpile by dripping sand grains randomly
with a more or less steady rate, until it forms a cone (Fig.~2, right).
Once the sandpile reaches a critical slope (typically at an
angle of repose of $\approx 34^\circ - 37^\circ$), avalanches occur
in an unpredictable way, even when the input of dripped sand grains
is steady. Bak called this driven state {\sl`` self-organized criticality''}
(SOC), which has become a paramount characteristic of many nonlinear systems.
A sand beach, in contrast, would be in a sub-critical state without 
avalanches (Fig.~2, left). The key features of nonlinear systems are
powerlaw-like size distributions, which indicate scale-free parameter distributions,
such as the size, time scale, or energy of an avalanche. The occurrence
frequency distribution $N(x)$ of a parameter $x$ is defined as a powerlaw 
function when 
$$
	N(x) \propto x^{-\alpha} \ . 
	\eqno(1)
$$
The mathematical nature of such powerlaw distributions can easily be
understood in terms of a coherent, exponential growth process as described
in the following section.

\begin{figure}
\centerline{\includegraphics[width=1.00\textwidth]{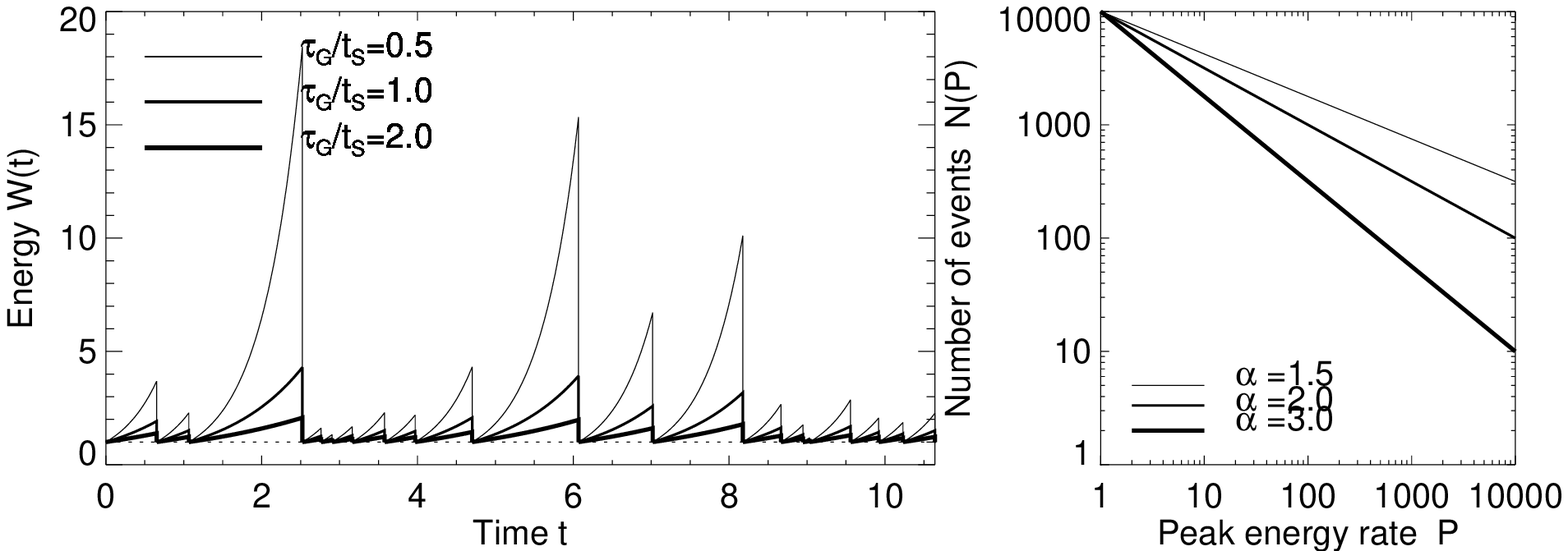}}
\captio{Time evolution of energy release rate $W(t)$ for 3
different ratios of growth times to saturation times,
$\tau_G/t_S=(0.5,1.0,2.0)$ (left) and the corresponding powerlaw
distributions of the peak energy release rate $P$. Note that the
event set with the shortest growth time ($\tau_G/t_S=0.5$)
reaches the highest energies and thus
produces the flattest powerlaw slope ($\alpha=1+\tau_G/t_S=1.5$).}
\end{figure}

\section{Analytical Model of SOC Avalanches} 

Avalanches occurring in the state of self-organized criticality represent
local instabilities that grow explosively for some time interval. The
released energy grows in a nonlinear way above some energy threshold,
which can be parameterized by some nonlinear function, for instance by
an exponential growth function. We define the time evolution of the
energy release rate $W(t)$ of a nonlinear process that starts at a
threshold energy of $W_0$ by
$$
        W(t) = W_0 \ \exp{\left({t \over {\tau}_G} \right)} \ ,
        \qquad 0 \le t \le \tau \ ,
	\eqno(2)
$$
where ${\tau}_G$ represents the exponential growth time. The process
grows exponentially until it saturates at time $t=\tau$ with a
saturation energy $W_S$,
$$
        W_S = W(t=\tau) = W_0 \ \exp{\left({\tau \over {\tau}_G} \right)} \ .
	\eqno(3)
$$
We define a peak energy release rate $P$ that represents the maximum
energy release rate $W_S$, after subtraction of the threshold energy $W_0$,
that corresponds to the steady-state energy level before the nonlinear
growth phase,
$$
        P = W_S - W_0 = W_0 \left[ \exp{ \left( {\tau \over \tau_G} \right)}
                - 1 \right] \ .
        \eqno(4)
$$
In the following, we will refer to the peak energy release rate $P$
also briefly as ``peak energy''. For the saturation times $\tau$,
which we also call ``rise times'', we assume a random probability distribution,
approximated by an exponential function $N(\tau)$ with e-folding time constant
$t_S$,
$$
        N(\tau ) d\tau = {N_0 \over t_S}
        \exp{\left(-{\tau \over t_S}\right)} d\tau \ .
        \eqno(5)
$$
This probability distribution is normalized to the total number of $N_0$
events,
$$
        \int_0^{\infty} N(\tau) d\tau = N_0 \ .
        \eqno(6)
$$
In order to derive the probability distribution $N(P)$ of peak energy
release rates $P$, we have to substitute the variable of the peak energy,
$P$, into the function of the rise time $\tau(P)$,
$$
        N(P) dP = N(\tau ) d\tau =
        N[\tau(P)] \left| {d\tau \over dP} \right| dP \ .
        \eqno(7)
$$
This requires the inversion of the evolution function
$P(\tau)$ (Eq.~4),
$$
        \tau(P) = \tau_G \ \ln{\left( {P \over W_0} + 1 \right)} \ ,
        \eqno(8)
$$
and the calculation of its derivative $d\tau/dP$, which is
$$
        {d\tau \over dP} = {\tau_G \over W_0}
         \left( {P \over W_0} + 1 \right)^{-1} \ .
        \eqno(9)
$$
Inserting the probability distribution of saturation times $N(\tau)$
(Eq.~5), the inverted evolution function $\tau(P)$ (Eq.~8) and
its time derivative $(d\tau/dP)$ from Eq.~(9) into the frequency
distribution $N(P)$ in Eq.~(7) yields then,
$$
        N(P) dP = {N_0 (\alpha - 1) \over W_0}
        \left({P \over W_0} + 1 \right)^{-\alpha} \ dP \ ,
        \eqno(10)
$$
which is an exact powerlaw distribution for large peak energies
($P \gg W_0$) with a powerlaw slope $\alpha$ of
$$
        \alpha = \left( 1 + {\tau_G \over t_S} \right) \ .
        \eqno(11)
$$
The same mathematical result was derived in Rosner and Vaiana (1978),
but the exponential growth was attributed to energy storage therein,
rather than to the nonlinear growth phase of the instability here.
The powerlaw slope thus depends on the ratio of the growth time to the
e-folding saturation time, which is essentially the average number of
growth times. Examples of time series with avalanches of different
growth times ($\tau_G/t_S=0.5, 1.0, 2.0$) are shown in Fig.~3,
along with the corresponding powerlaw distributions of peak energies $P$.
Note that the fastest growing events produce the flattest powerlaw
distribution of peak energies. 

This simple model explains the powerlaw distribution of peak energies
in nonlinear processes that have an exponential-like growth, which is
the key characteristic of coherent processes. In incoherent processes,
the growth time would be much larger than the average saturation time
($\tau_G \gg t_S$), in which case the time evolution (Eq.~3) can be
linearized, leading to $P \propto \tau$ (Eq.~4), 
and the distribution of peak energies
would be identical to that of the saturation times $\tau$, which
is a Poissonian (exponential) distribution (Eq.~5) with a characteristic
scale $P \approx W_0$. So, incoherent processes would have the same
exponential distribution of peak fluxes $P$ as their random durations $\tau$,
$$
        N^{inc}(P) dP = {N_0 \over W_0}
        \exp{\left(-{P \over W_0}\right)} dP \ .
        \eqno(12)
$$
Therefore, we can use the functional shape of the occurrence frequency
distribution of event sizes as a diagnostic of linear or nonlinear processes:
exponential distribution functions indicate an incoherent random process,
while powerlaw distribution functions indicate a coherent growth process.

\begin{figure}
\centerline{\includegraphics[width=1.00\textwidth]{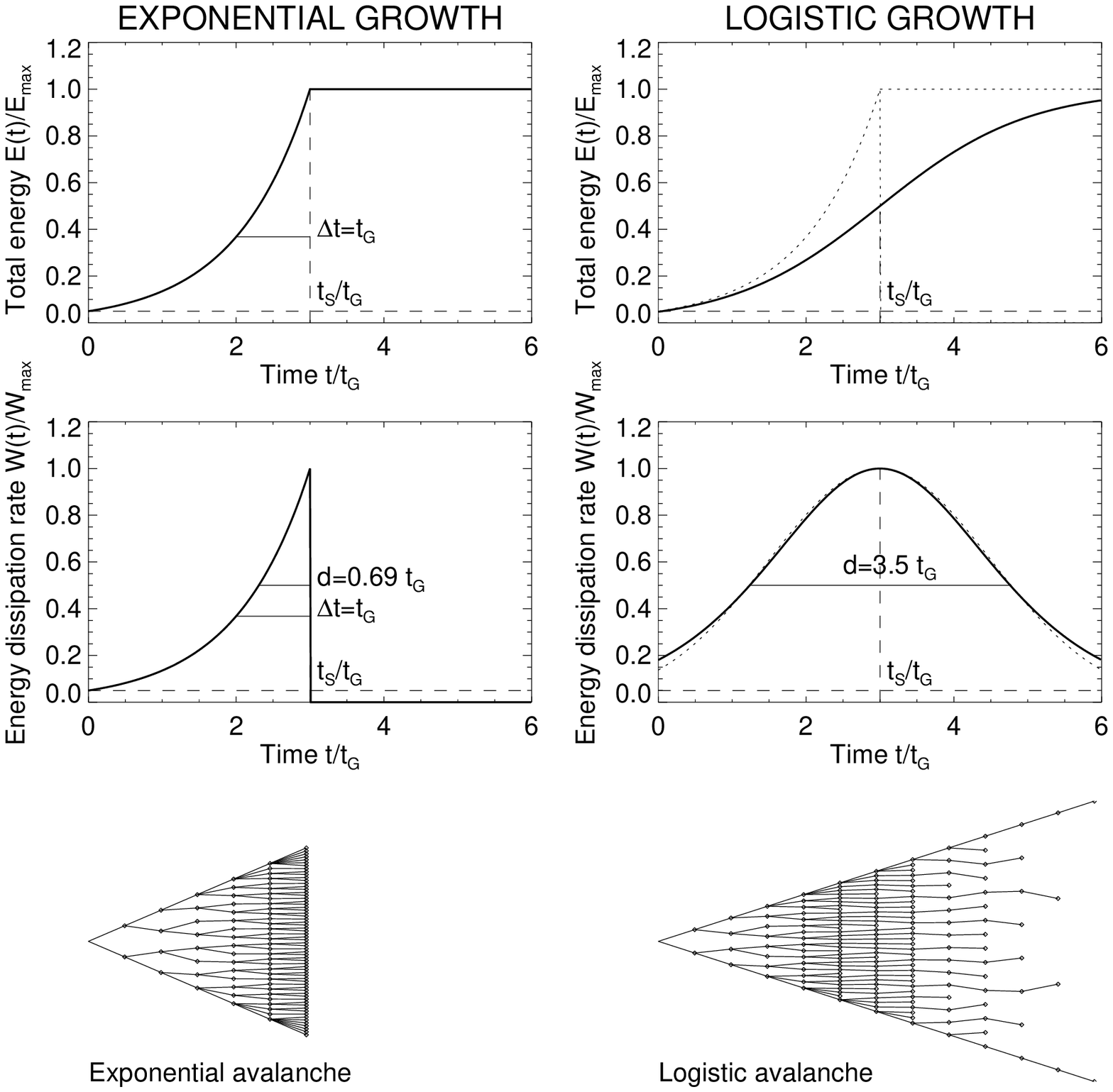}}
\captio{Time evolution of total released energy $e(t)$ (top
panels), the energy release rate $W(t)=de(t)/dt$ (middle
panels), and binary representation of avalanche growth rate
(bottom panels), for both the exponential (left panels) and the
logistic growth model (right panels). An exponential curve
(right top) and a Gaussian curve (right middle) are
drawn (with dotted lines) onto the logistic curves for comparison
(Aschwanden et al.~1998).}
\end{figure}

The exponential-growth model represents just the mathematically simplest
function to produce a powerlaw distribution of peak energies. It corresponds
to a multiplicative process where the number of avalanche elements doubles
every time step until it saturates (Fig.~4, left). The saturation phase,
however, abruptly ends with a discontinuity. A more natural model is
the logistic equation (discovered by Pierre Fran\c{c}ois Verhulst in 1845),
which is a first-order differential equation that approaches the saturation
limit asymptotically, 
$$
        {dE(t) \over dt } = {E(t) \over \tau_G} \cdot
        \left[1 - {E(t) \over E_\infty} \right]
        \eqno(13)
$$
as shown in Fig.~4 (right). The time derivative, which mimics the energy
release rate, has a maximum when the total energy has the steepest slope.
The avalanche grows first exponentially and dies out gradually after
the peak time of the fastest growth. It can be shown mathematically
that the frequency distribution of the peak energies is also a powerlaw
(Aschwanden et al.~1998), and thus this more physical description of a
{\sl logistic avalanche} with a continuous time evolution shares the same 
characteristic as an exponentially-growing avalanche, and thus both models
can be used to describe SOC processes. 

\begin{figure}
\centerline{\includegraphics[width=1.0\textwidth]{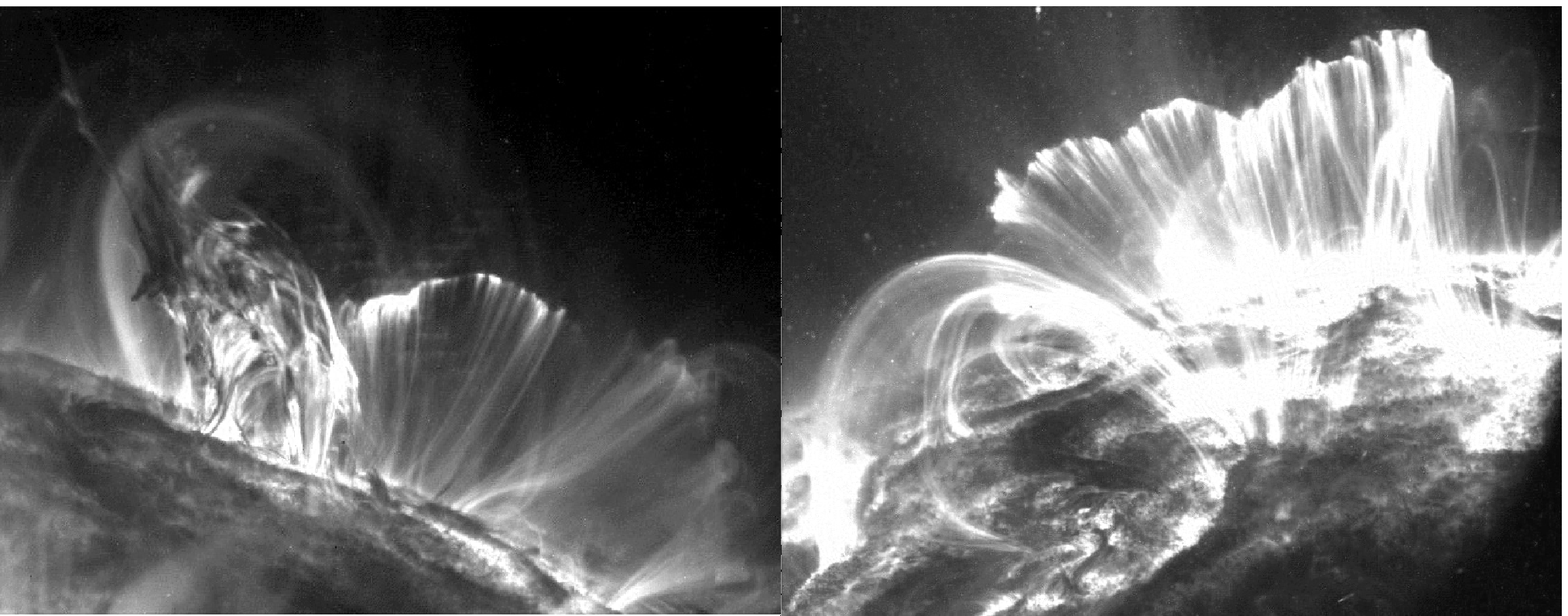}}
\captio{Solar flares observed in EUV with the TRACE spacecraft in 171 \ang :
The flare of 2001 Apr 15 exhibits an erupting filament in the foreground
and a rising postflare arcade behind near the limb (left panel),
while the 2000 Nov 9 flare displays the 3D geometry of the double-ribbon
postflare arcade (right panel) [courtesy of NASA/TRACE].}
\end{figure}

\begin{figure}
\centerline{\includegraphics[width=1.0\textwidth]{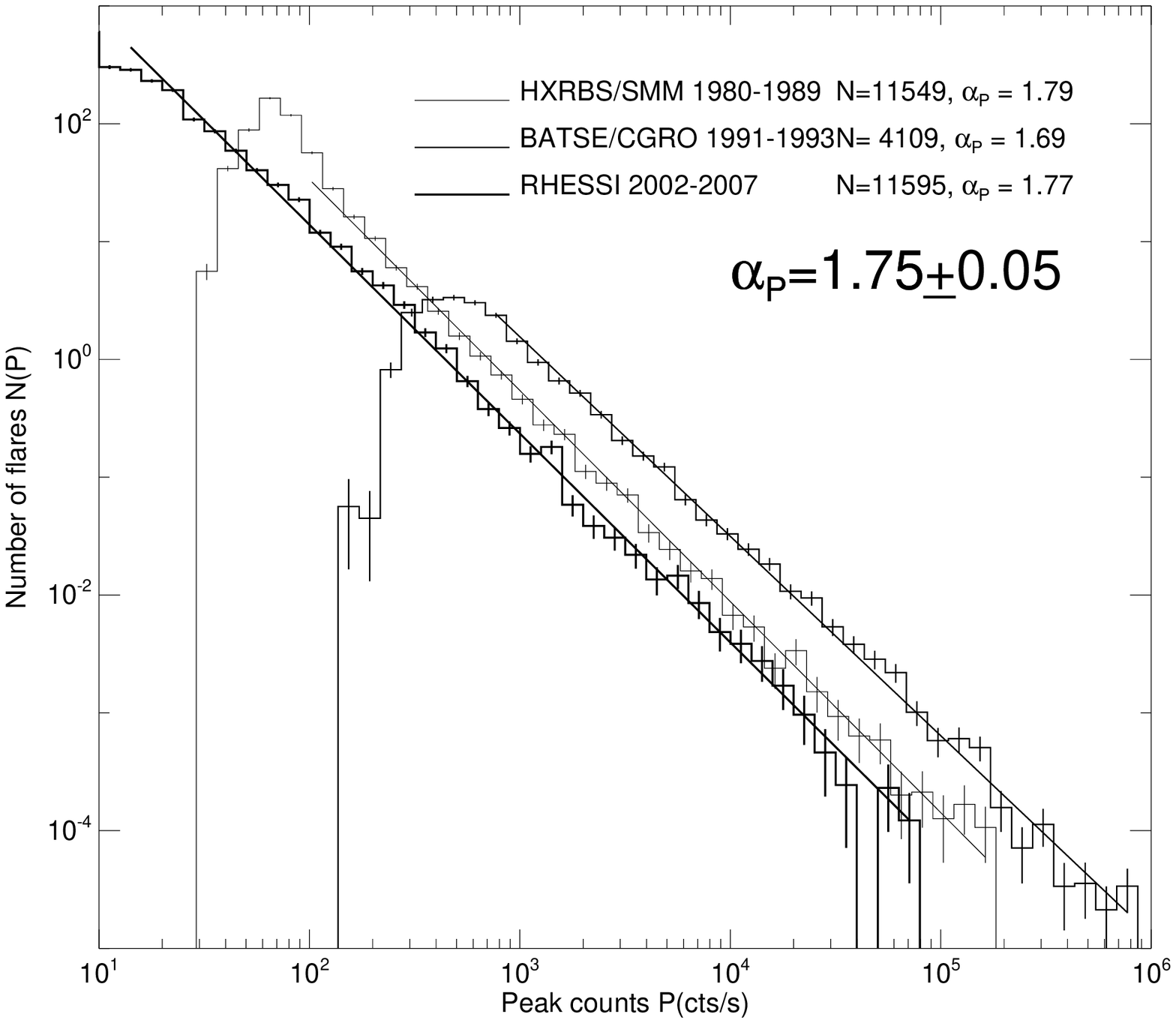}}
\captio{Occurrence frequency distributions of hard X-ray peak count rates
$P(cts/s)$ observed with HXRBS/SMM (1980-1989), BATSE (1991-1993),
and RHESSI (2002-2007), with powerlaw fits. Note that BATSE/CGRO has larger
detector areas, and thus records higher count rates.}
\end{figure}

\begin{figure}
\centerline{\includegraphics[width=0.9\textwidth]{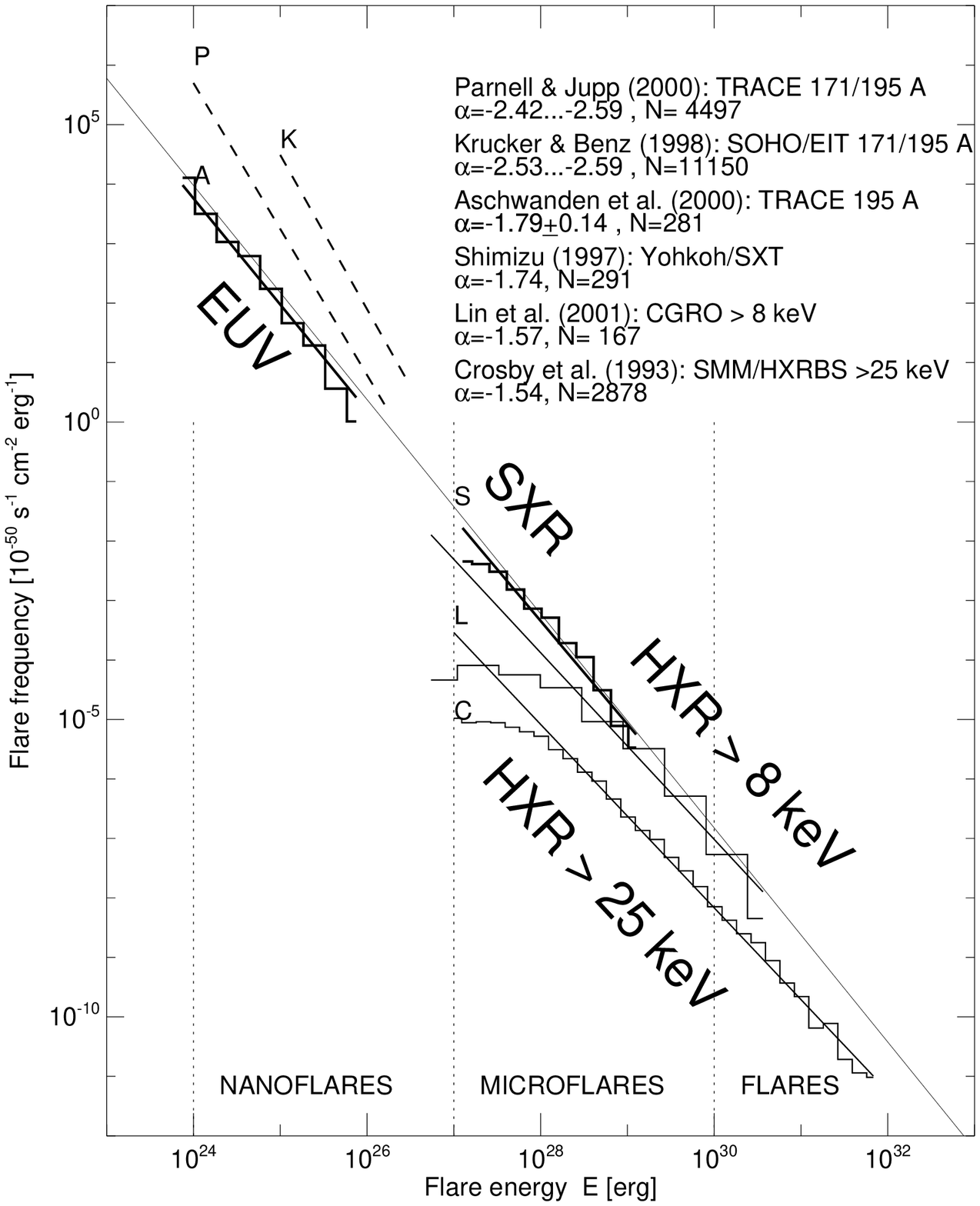}}
\captio{Composite flare frequency distribution in a normalized scale
in units of $10^{-50}$ flares per time unit ($s^{-1}$), area unit
(cm$^{-2}$), and energy unit (erg$^{-1}$). The energy is defined
in terms of thermal energy $E_{th}=3 n_e k_B T_e V$ for EUV and SXR,
and in terms of non-thermal energy in $>25$ keV (Crosby et al. 1993)
or $>8$ keV electrons (Lin et al.~2001). The slope of $-1.8$
is extended over the entire energy domain of $10^{24}-10^{32}$ erg.
The offset between the two HXR datasets is attributed to different
lower energy cutoffs as well as different levels of flare activity
during the observed time intervals [adapted from Aschwanden et al.~2000
and Lin et al.~2001].}
\end{figure}

\begin{figure}
\centerline{\includegraphics[width=0.8\textwidth]{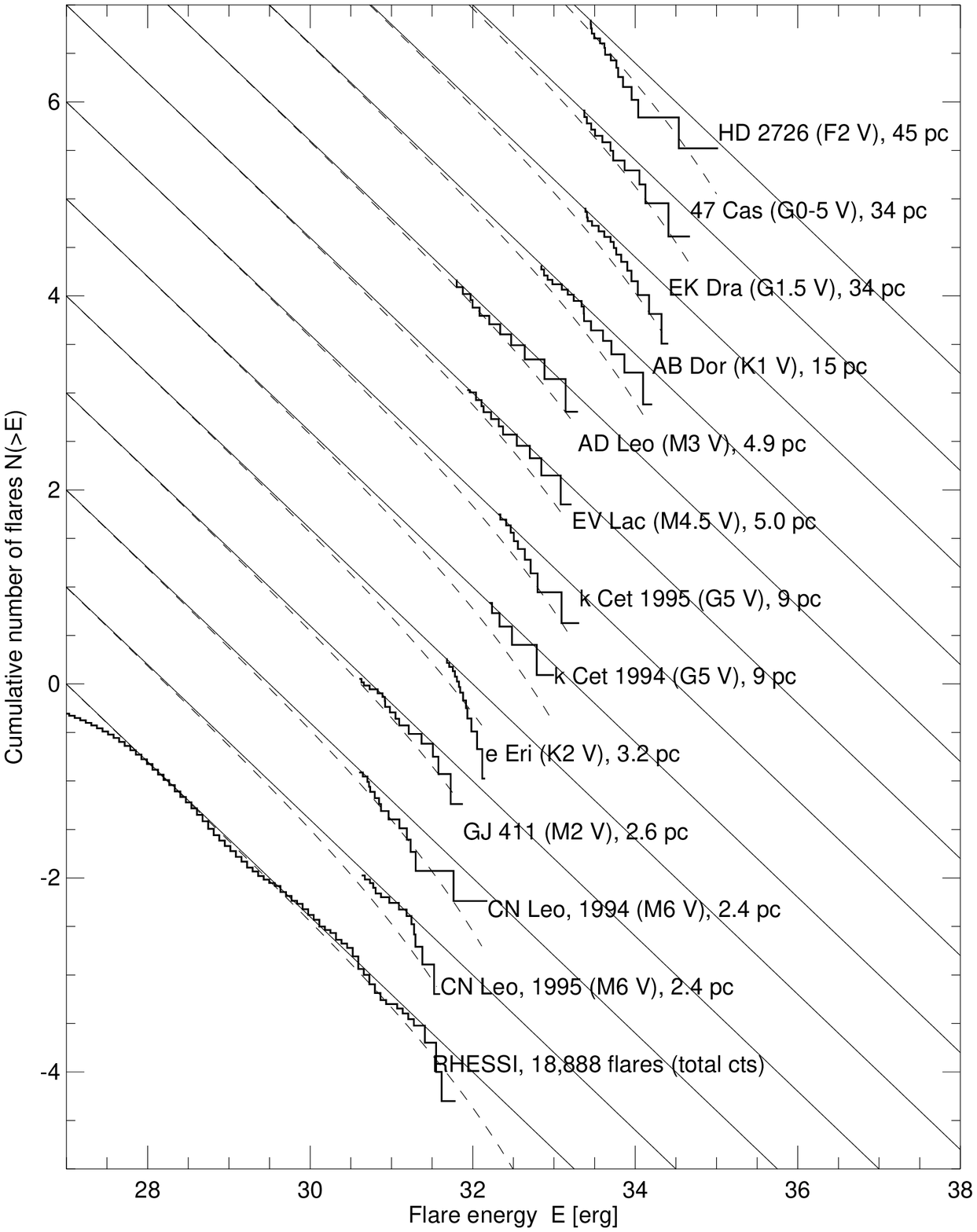}}
\captio{Cumulative occurrence frequency distributions $N(>E)$ of stellar
flares (vertically shifted for each star, in order of stellar distance)
are shown as a function of the flare energy $E$, observed from 12 different
stars by Audard et al.~(2000). For comparison we show also the statistics of
total hard X-ray emission of 18,888 solar flares observed with RHESSI,
which covers an energy range of $E\approx 10^{27.0}-10^{31.7}$ ergs, while
stellar flares have an energy range of $E\approx 10^{30.7}-10^{35.0}$ ergs.
Note that most distributions show a (cumulative) powerlaw part with a slope of
$\alpha \approx 0.8$ (diagonal lines) at the low end and an exponential
high-energy drop-off.}
\end{figure}

\section{Solar Observations of SOC Phenomena}

Solar flares are probably the best-studied datasets regarding SOC statistics
in astrophysics. Solar flares are catastrophic events in the solar corona,
most likely caused by a magnetic instability that triggers a magnetic
reconnection process, producing emission in almost all wavelengths,
such as in
gamma-rays, hard X-rays, soft X-rays, extreme ultraviolet (EUV), H$\alpha$
emission, radio wavelengths, and sometimes even in white light. 
Two EUV images of solar flares are shown in Fig.~5, which display the
intricated topology of the magnetic field produced at the beginning
(Fig.~5, left) and end phase of the flare (Fig.~5, right). Since the
emission mechanisms are all different in each wavelength, such as
nonthermal bremsstrahlung (in hard X-rays), thermal bremsstrahlung (in
soft X-rays and EUV), gyrosynchrotron emission (in microwaves), plasma emission
(in metric and decimetric waves), etc., we expect that the calculation of
energies contained in each event strongly depends on the emission mechanism,
and thus on the wavelength. It is therefore advisable to investigate the
statistics of SOC events in each wavelength domain separately. The most
unambiguous SOC parameters to report are the peak flux $P$, the total flux
or fluence $E$, defined as the time-integrated flux over the entire event,
and the total time duration $T$ of the event. Conversions of fluxes and
fluences into energy release rates and total energies require physical
models.

Statistics of the peak count rates $P$ of over 20,000 solar flares
observed in hard X-rays during three different space missions
(SMM 1980-1989; CGRO 1991-1993; RHESSI 2002-2010) and during three
different solar cycles have been gathered and their occurrence
frequency distributions (or log N - log S distributions) are shown
in Fig.~6. All three datasets show a near-perfect powerlaw distribution
function with a mean slope of $\alpha_P=1.75\pm0.05$. According to
our model this value corresponds to a ratio of $\tau_G/t_S=0.75$ (Eq.~11)
or a mean saturation time $t_S/\tau_G=1.33$ growth times. The distributions
shown in Fig.~6 span together over 5 orders of magnitude, which means
that the largest flare had a maximum saturation time of $\ln{(10^5)}=11.5$
growth times, which corresponds to an extremely coherent process that
could never be explained by random probability. 

Similar powerlaw-like occurrence frequency distributions have also been
found for other flare parameters, such as the fluence, the flare duration,
or flare energy (calculated from the thermal energy or energy in nonthermal
hard X-ray producing electrons). Based on this fact of scale-free powerlaw
distributions, Lu and Hamilton (1991) were the first to apply the concept
of self-organized criticality (SOC) to solar flares. With a cellular
automaton model that mimics sandpile avalanches, they were able to simulate 
the size frequency distribution with a powerlaw slope of $\alpha_P=1.67\pm0.02$,
which is close to the observed value of $\alpha_P=1.75\pm0.05$ shown in
Fig.~6. 

The statistics of peak energies in solar flares has been extended to
other wavelengths, from hard X-rays to soft X-rays and EUV and the
powerlaw-like size distribution has been found to extend over 8 orders
of magnitude (Fig.~7). The events observed in soft X-rays have been
called {\sl ``active region transients''}, having about 3-5 orders
of magnitude less energy than the largest solar flare. The events observed
in EUV have energies about 6-9 orders of magnitude less than the largest
flare, and thus are called {\sl ``nanoflares''}. The self-similarity of
flare parameters observed over this wide wavelength range, which implies
also a large temperature range of $T \approx (1-35) \times 10^6$ K, 
constitutes a powerful argument that the solar corona is in a state
of self-organized criticality.

\begin{figure}
\centerline{\includegraphics[width=0.5\textwidth]{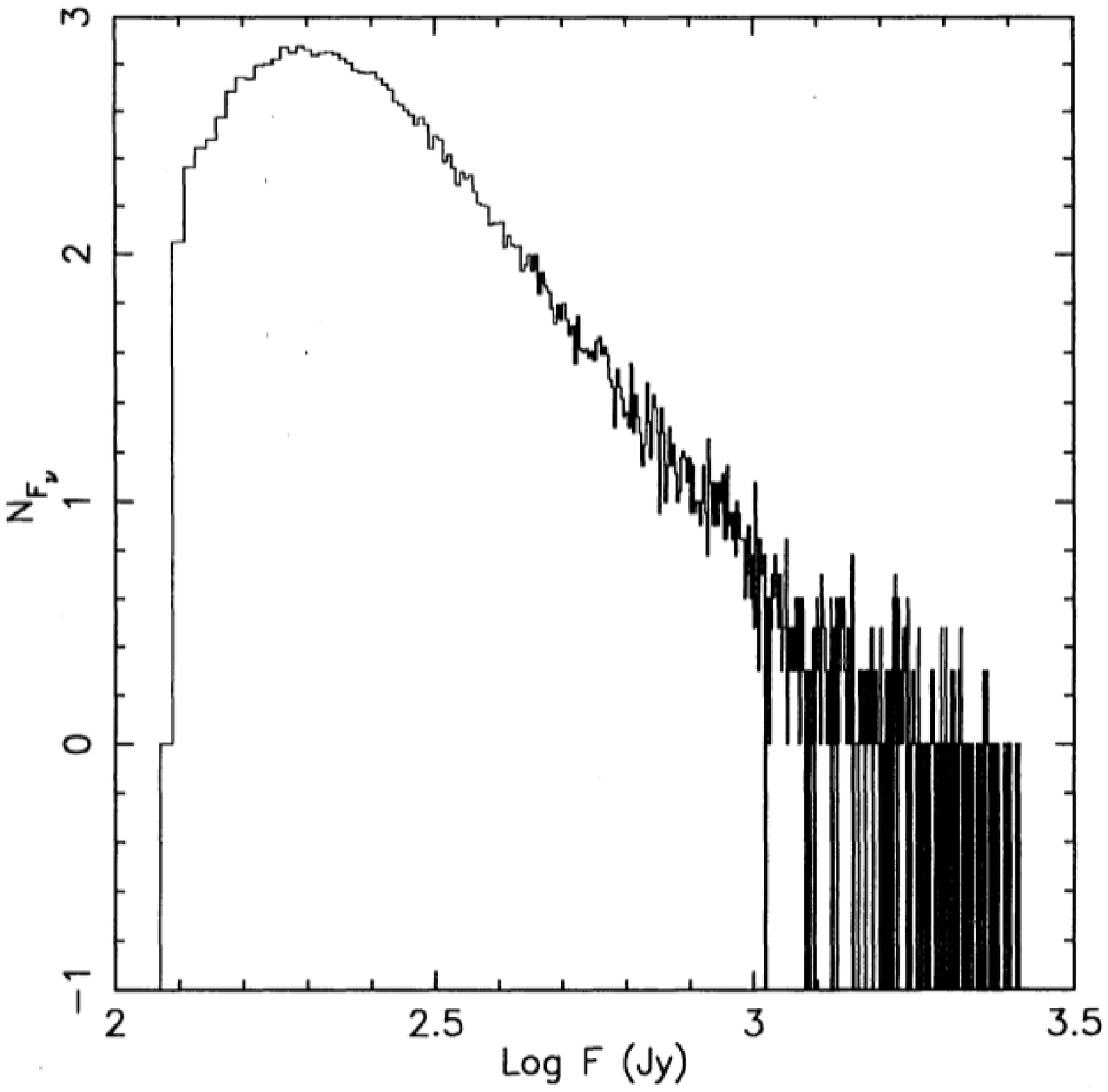}}
\captio{Frequency distribution of giant-pulse flux densities measured
from the Crab pulsar, observed during 15-27 May 1991 with the Green Bank
43-m telescope at 1330, 800, and 812.5 MHz. The tail can be represented
by a powerlaw distribution $N_F \propto F^{-\alpha}$ with $\alpha=3.46\pm0.04$
for fluxes $F>200$ Jy (Lundgen et al.~1995).}
\end{figure}

\section{Astrophysical Observations of SOC Phenomena}

Size distributions have also been sampled for stellar flares, although
with much smaller statistics, typically $\lapprox 15$ flare events per star.
A comparison of size distributions from 12 flare stars (type F to M) 
observed by Audard et al.~(2000) is shown in Fig.~8. Because of the
smallness of the samples, the cumulative frequency distributions are
shown, which have slope that is flatter by one compared with the
differential frequency distribution ($\alpha^{cum}_P = \alpha_P-1$).
The comparison in Fig.~8 demonstrates two important aspects: (1) 
stellar flares are up to two orders of magnitude more energetic
($E \approx 10^{32}-10^{34}$) than solar flares (see RHESSI comparison
at bottom of Fig.~8), although the small sample implies common
and relatively weak events, and (2) the drop-off at the upper bound of 
the cumulative frequency distribution yields a steeper slope than the
powerlaw slope obtained from more complete frequency distributions.
Correcting for the latter effect, stellar flares seem to have a similar
powerlaw distribution as solar flares, and thus may be goverend by the
same SOC process. 

\medskip
Other size distributions with powerlaw shape have been observed in
astrophysical data for pulsar glitches (e.g., Lundgren et al.~1995;
shown in Fig.~9), for soft gamma-ray repeaters (e.g., Gogus et al.~1999), 
accretion disk objects around black hole candidates, such as Cygnus X-1
(e.g., Negoro et al.~1995), or for blazars (e.g., Ciprini et al.~2003). 
These astrophysical objects display size distributions with different 
powerlaw values, and thus can be explained by different physical
processes, but the distribution type of a powerlaw function constitutes
a powerful argument that the underlying system is driven into the critical 
state of self-organized criticality, which is common to earthquakes,
sunquakes (occurring during large flares), and neutron star quakes 
(pulsar glitches).

\bigskip
{\sl Acknowledgements:
This work is partially supported by NASA contract
NAS5-98033 of the RHESSI mission through University of California,
Berkeley (subcontract SA2241-26308PG) and NASA grant NNX08AJ18G.
We acknowledge access to solar mission data and flare catalogs from the
{\sl Solar Data Analysis Center(SDAC)} at the NASA Goddard Space Flight
Center (GSFC).}

\section*{References} 

\footnotesize
\ref{Aschwanden, M.J., Dennis, B.R., and Benz, A.O. 1998,
        {\sl Logistic avalanche processes, elementary time structures,
        and frequency distributions of flares},
        Astrophys. J. {\bf 497}, 972-993.}
\ref{Aschwanden, M.J., Tarbell, T., Nightingale, R., Schrijver, C.J.,
        Title, A., Kankelborg, C.C., Martens, P.C.H., and Warren, H.P.
        2000, {\sl Time variability of the quiet Sun observed with TRACE:
        II. Physical parameters, temperature evolution, and energetics
        of EUV nanoflares}, Astrophys. J. {\bf 535}, 1047-1065.}
\ref{Audard, M., Guedel, M., Drake, J.J., and Kashyap, V.L. 2000,
        {\sl Extreme-ultraviolet flare activity in late-type stars}
        Astrophys. J. {\bf 541}, 396-409.}
\ref{Bak, P., Tang, C., \& Wiesenfeld, K. 1987,
        {\sl Self-organized criticality - An explanation of 1/f noise},
        Physical Review Lett. {\bf 59/27}, 381-384.}
\ref{Ciprini, S., Fiorucci, M., Tosti, G., and Marchili, N. 2003,
        {\sl The optical variability of the blazar GV 0109+224. Hints of
        self-organized criticality},
        in {\sl High energy blazar astronomy}, ASP Conf. Proc. {\bf 229},
        (eds. L.O. Takalo and E. Valtaoja), ASP: San Francisco, p.265.}
\ref{Crosby, N.B., Aschwanden, M.J., and Dennis, B.R. 1993,
 	{\sl Frequency distributions and correlations of solar hard X-ray 
	flare parameters}, Solar Phys. {\bf 143}, 275-299.}
\ref{Gogus, E., Woods, P.M., Kouveliotou, C., van Paradijs, J.,
        Briggs, M.S., Duncan, R.C., and Thompson, C. 1999,
        {\sl Statistical properties of SGR 1900+14 bursts},
        Astrophys. J. {\bf 526}, L93-L96.}
\ref{Krucker, S. and Benz, A.O. 1998,
 	{\sl Energy distribution of heating processes in the quiet solar 
	corona}, Astrophys. J. {\bf 501}, L213-L216.}
\ref{Lin, R.P., Feffer,P.T., and Schwartz,R.A. 2001,
        {\sl Solar Hard X-Ray Bursts and Electron Acceleration Down to 8 keV},
        Astrophys. J. {\bf 557}, L125-L128.}
\ref{Lu, E.T. and Hamilton, R.J. 1991,
        {\sl Avalanches and the distribution of solar flares},
        Astrophys. J. {\bf 380}, L89-L92.}
\ref{Lundgren, S.C., Cordes,J.M., Ulmer, M., Matz, S.M., Lomatch,S.,
        Foster, R.S., and Hankins,T. 1995,
        {\sl Giant pulses from the Crab pulsar: A joint radio and
        gamma-ray study}, Astrophys. J. {\bf 453}, 433-445.}
\ref{Negoro, H., Kitamoto, S., Takeuchi, M., and Mineshige, S. 1995,
        {\sl Statistics of X-ray fluctuations from Cygnus X-1:
        Reservoirs in the disk ?}, Astrophys. J. {\bf 452}, L49-L52.}
\ref{Parnell, C.E. and Jupp, P.E. 2000,
 	{\sl Statistical analysis of the energy distribution of nanoflares 
	in the quiet Sun}, Astrophys. J. {\bf 529}, 554-569.}
\ref{Rosner,R., and Vaiana,G.S. 1978,
	{\sl Cosmic flare transients: constraints upon models for energy 
	storage and release derived from the event frequency distribution},
 	Astrophys. J. {\bf 222}, 1104-1108.}
\ref{Shimizu, T. and Tsuneta, S. 1997,
 	{\sl Deep survey of solar nano-flares with Yohkoh},
 	Astrophys. J. {\bf 486}, 1045-1057.}

\end{document}